\documentclass[amsmath,amssymb,
showpacs,twocolumn, nobibnotes,aps,prb]{revtex4-1}

\usepackage{graphicx}
\usepackage{dcolumn}
\usepackage{bm}
\usepackage{enumerate}
\usepackage{amsmath,amssymb,amsfonts}
\usepackage{verbatim}
\usepackage{color}
\usepackage{mathrsfs}
\usepackage{setspace}
\usepackage{mathtools}
\usepackage{braket}
\usepackage{lipsum}

\newcommand{\be}{\begin{equation}}
\newcommand{\ee}{\end{equation}}
\def\ba{\begin{aligned}}
\def\ea{\end{aligned}}

\newcommand{\bea}{\begin{eqnarray}}
\newcommand{\eea}{\end{eqnarray}}

\renewcommand{\hat}[1]{{\widehat #1}}

\begin{document}

\title{Backscattering of topologically protected helical edge states by line defects}

\author{Mohadese Karimi}
\author{Mohsen Amini}
\email{msn.amini@sci.ui.ac.ir}
\author{Morteza Soltani}  
\email{mo.soltani@sci.ui.ac.ir}
\author{Mozhgan Sadeghizadeh}

\affiliation{Department of Physics, Faculty of Physics, University
of Isfahan, Isfahan 81746-73441, Iran}

\begin{abstract}
The quantization of conductance in the presence of non-magnetic point defects is a consequence of topological protection and the spin-momentum locking of helical edge states in two-dimensional topological insulators. This protection ensures the absence of backscattering of helical edge modes in the quantum Hall phase of the system. 
However, our study focuses on exploring a novel approach to disrupt this protection. We propose that a linear arrangement of on-site impurities can effectively lift the topological protection of edge states in the Kane-Mele model.
To investigate this phenomenon, we consider an armchair ribbon containing a line defect spanning its width. Utilizing the tight-binding model and non-equilibrium Green's function method, we calculate the transmission coefficient of the system. Our results reveal a suppression of conductance at energies near the lower edge of the bulk gap for positive on-site potentials.
To further comprehend this behavior, we perform analytical calculations and discuss the formation of an impurity channel. This channel arises due to the overlap of in-gap bound states, linking the bottom edge of the ribbon to its top edge, consequently facilitating backscattering. Our explanation is supported by the analysis of the local density of states at sites near the position of impurities.

\end{abstract}
\maketitle
\section{Introduction~\label{Sec01}}

The emergence of two-dimensional topological insulators (2D TIs) has revolutionized the field of condensed matter physics by introducing a fascinating phenomenon known as bulk-boundary correspondence~\cite{topol1}. This principle dictates the existence of metallic gap states that are confined to the one-dimensional edges of the material. These edge states exhibit a distinct property of spin polarization, where the spin of an electron becomes rigidly coupled to its momentum. 
The intriguing consequence of this phenomenon is that electrons traveling along the edge exhibit remarkable immunity to backscattering caused by non-magnetic defects, leading to conductance quantization.
This phenomenon is known as topological edge state protection and lies at the heart of the celebrated quantum spin Hall effect~\cite{topol1, topol2,rev2020}. 

The study of topological protection in 2D TIs raises a crucial question concerning the mechanisms that can lift this protection. This issue holds a dual significance. 
Firstly, it pertains to the puzzling deviations from conductance quantization observed in experiments on 2D TIs~\cite{PZ1, PZ2, PZ3}, necessitating a deeper understanding of the underlying mechanisms that induce the lifting of protection~\cite{US1, US2, US3}. Secondly, in many practical scenarios, the ability to control and manipulate edge states becomes paramount~\cite{Yazdani, APP1, APP2, APP3, APP4, APP5, APP6}. Depending on the specific device and application, there arises a need to modulate or suppress these edge states, underscoring the importance of comprehending the mechanisms behind the lifting of protection.

So far, numerous proposals have been put forth to explain the mechanisms behind the lifting of topological protection and the subsequent occurrence of backscattering in 2D TIs. 
Possible sources of backscattering include mechanisms that violate time-reversal symmetry, such as the influence of an external magnetic field~\cite{MF}, the presence of charge puddles~\cite{PRL2013}, the interaction with embedded nuclear spins~\cite{SP1, SP2}, coupling to phonons~\cite{PRL2012}, and the impact of electromagnetic noise~\cite{PRL2018}.
However, a more direct and controllable approach for lifting topological protection is achieved through tunneling between opposite edges of a 2D TI~\cite{APP4, APP5, APP6, T1, T2, T3, T4}. 
The tunneling between opposite edges of a 2D TI enables the coupling of electrons moving in one direction with their counterparts of the same spin orientation on the opposing edge.
As a result, a small gap is opened at the Fermi level~\cite{Amini1}, and more significantly, a channel for electron backscattering is created without the need to break time-reversal symmetry. A potential scenario for achieving this condition involves the creation of spatially extended defects within a 2D TI, which can be engineered through methods such as nano-patterning or the introduction of specific line defects in the atomic lattice~\cite{Nan1}. 

Recently, a promising achievement in the observation of edge coupling  due to the presence of an extended linear defect in a 2D TI is reported experimentally~\cite{NP}.
In this experimental study, bismuthene, which serves as a prototypical 2D TI is utilized. Interestingly, it is observed that narrow constrictions spontaneously manifest themselves within the material in the form of line defects.
By employing scanning tunneling microscopy/spectroscopy (STM/STS) measurements and analyzing the local density of states (LDOS), it has been discovered that the presence of the line defect leads to a spatial overlap of the edge states localized along both edges of the line defect. This spatial overlap induces a phenomenon known as hybridization, which involves the mixing of the edge states from each edge\cite{NP}. As a consequence of this hybridization, inter-edge scattering occurs, giving rise to a gap in the energy spectrum. The presence of this gap creates a channel through which back scattering can take place.

In this study, we introduce a novel mechanism for the disruption of conductance quantization in  2D TIs. Our proposed mechanism involves the application of line defects, which result in the breakdown of conductance quantization without the need for direct coupling of edge states. Importantly, this breakdown occurs without the opening of a gap in the energy spectrum.
In this study, we focus on a 2DTI ribbon that contains a line of on-site impurities (line defect) along its longitudinal direction. Our objective is to investigate the robustness of the transport properties of the system using the non-equilibrium Green's functions formalism\cite{Datta}. By employing this framework, we aim to gain insights into how the presence of line defects affects the transport characteristics of the 2DTI ribbon.
Utilizing a combination of numerical simulations and analytical calculations, we have discovered that the strength of the on-site potential plays a crucial role in the emergence of quasi one-dimensional eigenstates along the line defect in the 2DTI ribbon. This intriguing phenomenon transforms the line defect into a quantum wire that facilitates the transportation of current across the width of the ribbon. Moreover, the presence of the line defect introduces a mechanism for backscattering, which affects the transport properties of the system. Our findings shed light on the intricate interplay between on-site potential, line defects, and the transport behavior in 2DTI ribbons, offering new insights into the manipulation and control of quantum states in these systems.

The rest of the paper is organized as follows.
In Section~\ref{Sec.II}, we provide a comprehensive description of the model system under consideration and outline the methodology employed to study the transmission through this system.
Moving to Section~\ref{Sec.III}, we present the results obtained from our quantum transport study of the system in the presence of the line defect. We thoroughly discuss the mechanism responsible for the breakdown of topological protection through analytical calculations, elucidating the formation of the impurity channel resulting from bound states in the bulk gap region.
Finally, in Section~\ref{Sec.IV}, we conclude our findings.

\section{Model and method}\label{Sec.II}

\begin{figure}[t!]
  	\includegraphics[width=1.05\linewidth]{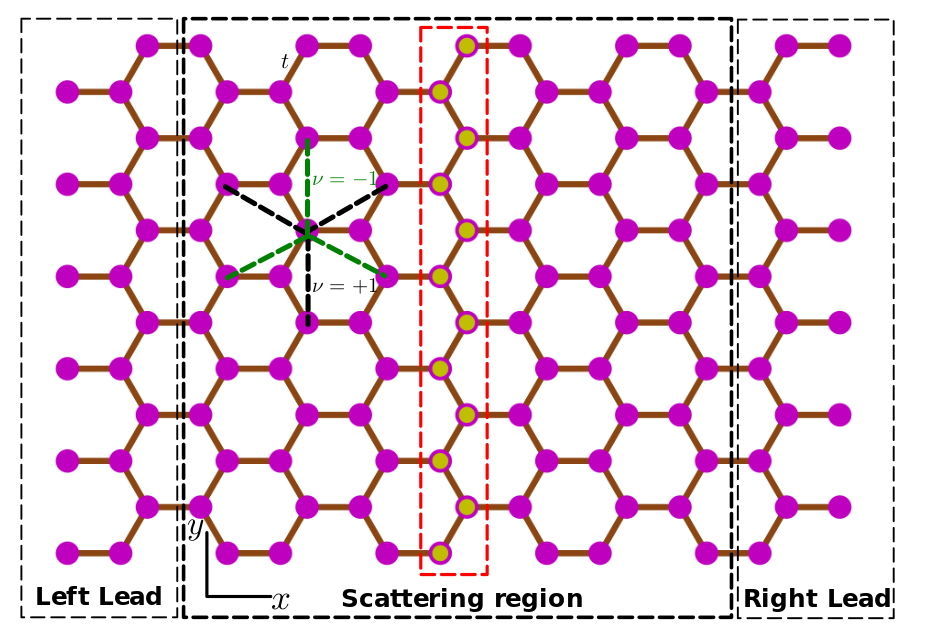}
	\caption{\label{F1}
	Schematic representation of a two-terminal device with an infinitely long Kane-Mele ribbon with armchair boundaries connected to macroscopic leads on the left and right sides. The scattering region contains a line defect of on-site impurities arranged in a zigzag chain across the width of the ribbon (yellow circles surrounded by dashed orange rectangle). The sign of $\nu_{ij}$ for the spin-orbit coupling term is shown in the scattering region. The width of the ribbon is $W=6$ armchair chains.}
\end{figure}

\subsection{Model}
In this section, we present a setup designed to investigate the transport properties of a 2D TI ribbon with an embedded line defect. Our device consists of two conducting leads that enclose a scattering region, as depicted in Fig. 1. The scattering region includes a line of sites with defects, which are visually represented by distinct colors in Fig.~\ref{F1}.
As it is shown we consider a two-dimensional honeycomb lattice with armchair edges along the $x$-direction.
In order to emulate the transport through an infinitely long ribbon in the $x$-direction, the ribbon under consideration in this model has a finite width $W$ in the $y$-direction. To achieve this, two semi-infinite leads are connected to the central region, as illustrated in Fig.~\ref{F1}.
The Hamiltonian governing the scattering region as well as the conducting leads in the tight-binding approximation is expressed as~\cite{KM1,KM2}:
\begin{eqnarray}\label{E1}
\label{KMMod}
\mathcal{H}_{\text{KM}} &=&t\sum_{\langle i,j\rangle,\alpha }c_{i\alpha}^{\dagger
}c_{j\alpha}+i\lambda _{\text{SO}}\sum_{\langle \langle i,j\rangle \rangle
,\alpha \beta}v_{ij}c_{i\alpha}^{\dagger }s^z_{\alpha \beta}c_{j\beta}   \\ 
&+&\lambda_\nu\sum_{i\alpha}\xi _{i}c_{i\alpha}^{\dagger
}c_{i\alpha}+ \text{h.c.}  \notag 
\end{eqnarray}%
This Hamiltonian consists of several terms. The first term corresponds to the nearest-neighbor (NN) hopping with amplitude $t$, where $c_{i\alpha}^\dagger (c_{i\alpha})$ denotes the creation (annihilation) operator for an electron with spin $\alpha$ at site $i$. The summation $\langle i, j \rangle$ runs over all the NN sites. 
The second term represents the intrinsic spin-orbit coupling (SOC) with coupling strength $\lambda_{\text{SO}}$ between next-nearest-neighbor (NNN) sites. The summation $\langle\langle i, j \rangle\rangle$ in the index indicates that it runs over all pairs of NNN sites and the Pauli matrices $\bm{s} = (s^x, s^y, s^z)$ are associated with the physical spins.
The factor $v_{ij}=\frac{\bm{d}_{i}\times \bm{d}_{j}}{|\bm{d}_{i}\times \bm{d}_{j}|}=\pm 1$ in the second term depends on the hopping path between NNN sites $i$ and $j$. It is determined by the cross product of the vectors $\bm{d}_{i}$ and $\bm{d}_{j}$ connecting the NNN sites, and takes values of $\pm 1$ as shown in Fig.~\ref{F1}.
The third term in the Hamiltonian corresponds to the staggered sublattice potential with strength $\lambda_{\nu}$. It introduces a potential difference between the two sublattices of the honeycomb lattice. The sublattice index $\xi_i$ takes values of $\pm 1$ depending on the sublattice of site $i$.
The term $h\cdot c$ represents the Hermitian conjugate of the previous terms.

To introduce a line of impurities, we incorporate a zigzag chain of sites spanning across the width of the ribbon in the scattering region. These sites possess an additional on-site potential $U$, creating a line defect indicated by a red rectangle in Fig.~\ref{F1}. The corresponding Hamiltonian for this line defect can be written as follows:
\be\label{E2}
\mathcal{H}_{\text{LD}} =  U \sum_{i \in \text{LD} ,\alpha}c_{i\alpha}^{\dagger
}c_{i\alpha}.
\ee

\subsection{Method}
We employ the non-equilibrium Green's functions formalism to investigate the electronic transport properties of our system~\cite{Datta}. This approach allows us to analyze the flow of electrons and calculate quantities such as the transmission coefficient and conductance.
The transmission coefficient ($\tau$) at zero temperature within this formalism can be computed using the Landauer-Buttiker formula, given by:
\be
\tau(E) = \text{Tr}[\Gamma_L(E)G_r(E)\Gamma_R(E)G_a(E)].
\ee
In this context, the retarded Green's function in the site representation, denoted as $G_r(E)$, is given by $G_r(E)=[E-\mathcal{H}_C-\Sigma_R(E)-\Sigma_L(E)]^{-1}$, where $\mathcal{H}_C=\mathcal{H}_{KM}+\mathcal{H}_{LD}$ represents the Hamiltonian of the scattering region, which incorporates the influence of the line defect. Similarly, the advanced Green's function, denoted as $G_a(E)$, is defined as the Hermitian conjugate of $G_r(E)$, i.e., $G_a(E)=[G_r(E)]^\dagger$. The self-energy terms $\Sigma_{R(L)}$ correspond to the embedding self-energy, which relies on the retarded contact Green's functions and the coupling between the leads $R(L)$ and the central (scattering) region.
The right and left line-width functions, denoted as $\Gamma_{R(L)}(E)$, can be expressed as $\Gamma_{R(L)}(E) = i[\Sigma_{R(L)} - (\Sigma_{R(L)})^\dagger]$.

\section{Results and discussion}\label{Sec.III}

\begin{figure}[t!]
  	\includegraphics[width=1.15\linewidth]{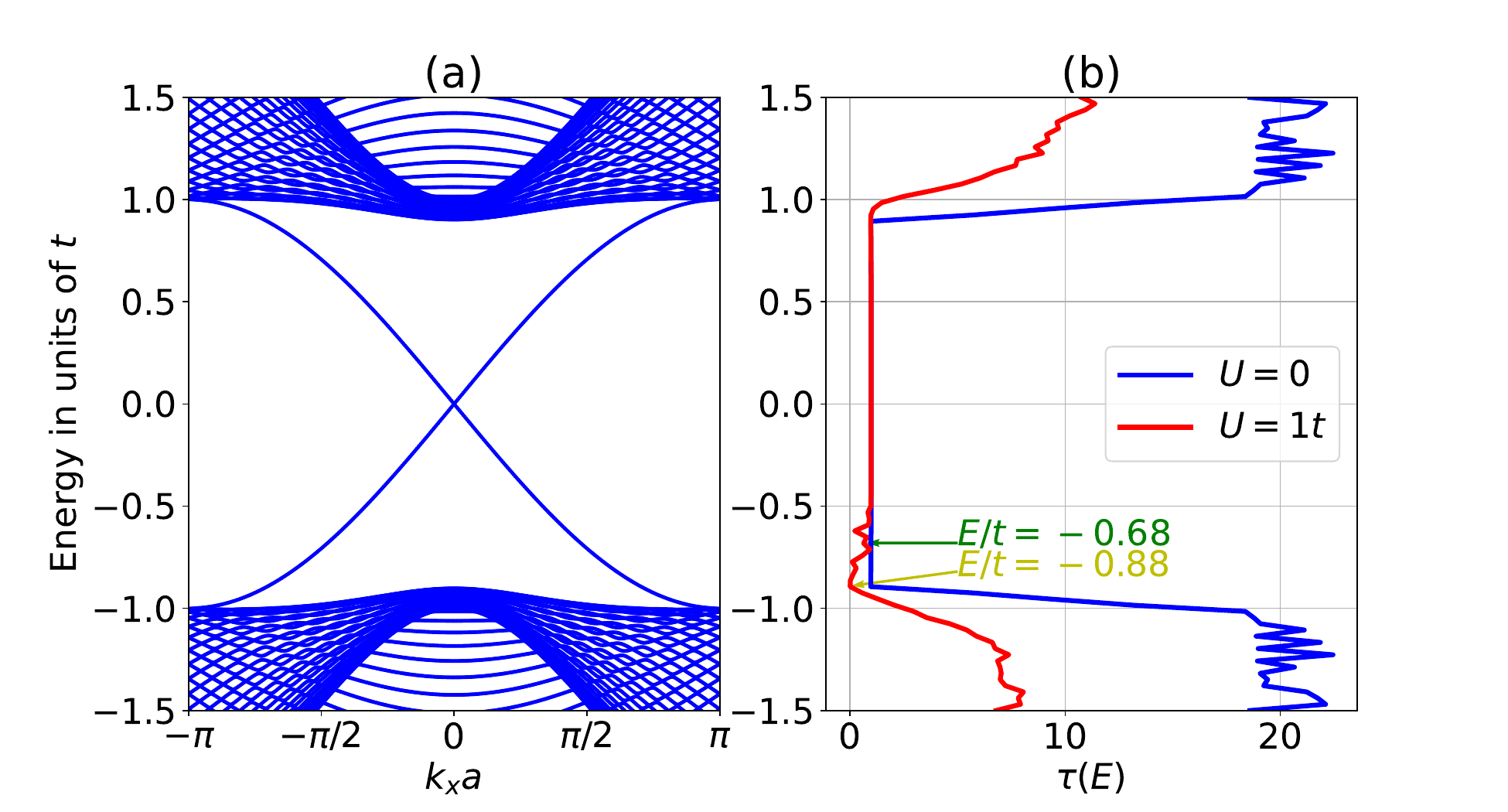}
	\caption{\label{F2}
(a) Electronic band structure of the pristine Kane-Mele model on a ribbon with armchair terminations and a width of $W=30$. The spin-orbit coupling strength is $\lambda_{\text{SO}}=0.2t$.
(b) The corresponding zero-temperature transmission coefficient $\tau(E)$ as a function of energy for both absence $(U=0)$ and presence $(U=1.0)$ of the line defect in the scattering region. 
The transmission coefficient curve demonstrates the quantization plateaus in the absence of the line defect, indicating topological protection, and the suppressed transmission regions caused by the line defect, leading to dips in the transmission coefficient curve. Two specific energy points with large and small suppression are highlighted at $E=-0.88t$ and $E=-0.68t$, respectively.
	}
\end{figure}

In this section, we analyze the electronic transport properties of the structure mentioned above. In our calculations, we take the NN hopping amplitude $t$ as the energy unit. Additionally, we set the width of the ribbon to be $W=30$, a value chosen to be sufficiently large to avoid coupling of edge states~\cite{Amini1} and fix $\lambda_{\text{SO}}=0.2t$.
To numerically calculate the transmission coefficients $\tau(E)$, we will utilize the PYQULA library~\cite{Pyqula}.

\subsection{Breakdown of the conductance  quantization}

 To begin, let us examine the band structure of the pristine system ($U=0$), which is illustrated in Fig.~\ref{F2}(a). As expected, the band structure displays a bulk band gap of approximately $\Delta=6\sqrt{3} \lambda_{\text{SO}}$~\cite{KM1, KM2}, along with the presence of edge bands near the gap region. The corresponding transmission coefficient $\tau(E)$ for the edge states is depicted by the blue lines in Fig.~\ref{F2}(b). Notably, in the vicinity of the gap region, the transmission coefficient of the edge states is $\tau(E)=1$, indicating their protected nature.

However, when we introduce a line defect with $U\neq 0$, the situation becomes significantly different. Our calculations reveal that for $U=1.0t$, the transmission coefficient of the edge states is suppressed for energies near the bottom of the bulk bands ($E\sim -0.9t$). This leads to the backscattering of the corresponding edge states and consequently results in the lifting of their protection. In the subsequent analysis, we will delve into the underlying reasons for this intriguing phenomenon.

\subsection{Analysis of the impurity channel formation}
In this subsection, we will demonstrate that the lifting of the topological protection of the edge states in the presence of the line defect can be attributed to the formation of an additional channel that connects the top edge of the ribbon to the bottom edge. To address this phenomenon analytically, we will examine the in-gap impurity states, which play a crucial role in the creation of this conducting channel.
To achieve this, we will begin by analyzing the in-gap impurity states of a Su-Schrieffer-Heeger (SSH)~\cite{SSH} model. Subsequently, we will extend and generalize these findings to our Kane-Mele ribbon. By studying the SSH model first, we can gain insights into the formation of in-gap impurity states, which will provide a foundation for our analysis of the Kane-Mele ribbon's behavior in the presence of a line defect.

\subsubsection{In-gap impurity state in the SSH model}

\begin{figure}[t!]
  	\includegraphics[width=1.0\linewidth]{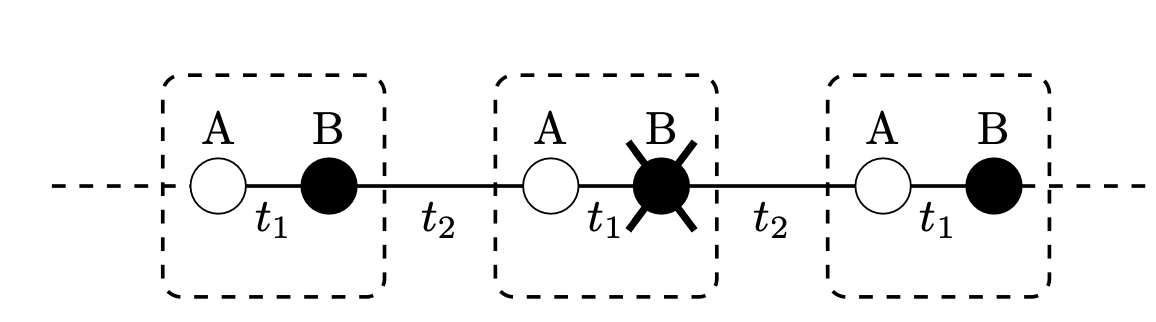}
	\caption{\label{F3}
	The geometry of an SSH chain with two types of sub-lattices, labeled as $A$ and $B$. The chain's unit cell is enclosed by a dashed rectangle, and the intracell and intercell hopping amplitudes are represented by $t_1$ and $t_2$, respectively.	}
\end{figure}

In this section, we investigate an SSH chain with an on-site impurity, as depicted in Fig.~\ref{F3}. The SSH chain consists of two nonequivalent sites, labeled as $A$ and $B$, within each periodic cell.
The Hamiltonian of the system is given by the sum of two terms, the SSH term and the impurity term, expressed as:
\begin{equation}
\mathcal{H} = \mathcal{H}_{\text{SSH}} + \mathcal{H}_{\text{impurity}},
\end{equation}
where
\begin{equation}
\label{E5}
\mathcal{H}_{\text{SSH}} = \sum_i t_1 a_i^\dagger b_i + t_2 b_i^\dagger a_{i+1} + \text{h.c.},
\end{equation}
and
\begin{equation}
\mathcal{H}_{\text{impurity}} = U b_0^\dagger b_0.
\end{equation}

Here, $a_i^\dagger (a_{i+1})$ and $b_i^\dagger (b_i)$ are the creation (annihilation) operators at the corresponding sites in the unit cell, with different coupling amplitudes denoted as $t_1$ and $t_2$ for the intra-cell and inter-cell couplings respectively.  The parameter $U$ represents the strength of the on-site impurity potential which is considered to be located at site $i=0$.

Let us begin by considering the case of the clean system, where $U=0$. 
We are primarily interested in the bulk properties of the system which implies  considering  a sufficiently  long chain~\cite{Book}. 
To analyze the bulk properties of the system in momentum space, we apply the Fourier transformation to the SSH term $\mathcal{H}_{\text{SSH}}$ of Eq.~\eqref{E5}, leading to the corresponding momentum-space Hamiltonian:
\begin{equation}
\mathcal{H}_{\text{SSH}}(k) = \sum_k \begin{bmatrix} a_k^\dagger & b_k^\dagger \end{bmatrix} \begin{bmatrix} h(k) \end{bmatrix} \begin{bmatrix} a_k \\ b_k \end{bmatrix},
\end{equation}
where
\begin{eqnarray}
h(k) &=& \begin{bmatrix} 0 & t_1 + t_2e^{ik} \\ t_1 + t_2e^{-ik} & 0 \end{bmatrix} \nonumber \\
&=& (t_1 + t_2\cos(k))\sigma_x + (t_2\sin(k)) \sigma_y.
\end{eqnarray}
In the above equation, $\sigma_x$ and $\sigma_y$ represent the Pauli matrices.
Now, the energy spectrum of $h(k)$ can be obtained by calculating its eigenvalues, which are given by:
\begin{equation}
E^\pm(k) = \pm \sqrt{t_1^2 + t_2^2 + 2t_1t_2\cos k}.
\end{equation}
As a result, the energy gap between the two energy bands is given by:
\begin{equation}
E_g =2|t_2 - t_1|.
\end{equation}

In this model, when the intercell hopping parameter $(t_1)$ is smaller than the intracell hopping parameter $(t_2)$, the chain exhibits a dimerized pattern, leading to a topologically nontrivial phase characterized by a winding number $(W)$ of 1~\cite{Book}. Conversely, when $t_1$ is larger than $t_2$, the chain becomes uniform, resulting in a trivial phase with $W=0$.
We will now focus on the topologically nontrivial phase where $t_2 > t_1$ and attempt to find the end modes, also known as edge states, which satisfy the Schrödinger equation $\mathcal{H}_{\text{SSH}}|\psi_{\text{edge}}\rangle=0$.
Due to the chiral symmetry of the system, the corresponding edge state can be expressed in such a way that it has support on only one sub-lattice~\cite{Book}, for example, sub-lattice $A$.
Therefore, we have $\langle 0 | b_i | \psi_{\text{edge}} \rangle = 0$ and the solution of the zero-energy eigenstate can be expressed as:
\begin{equation}
|\psi_{\text{edge}}\rangle = \sum_i \psi^A_i a_i^\dag |0 \rangle = \sum_i \left(-\frac{t_1}{t_2}\right)^i \psi^A_1 a_i^\dag|0\rangle,
\end{equation}
where the coefficient $\psi^A_1$ can be determined using the normalization condition as $(\psi^A_1)^2 = 1 - \left(\frac{t_1}{t_2}\right)^2$.

\begin{figure}[t!]
  	\includegraphics[width=1.0\linewidth]{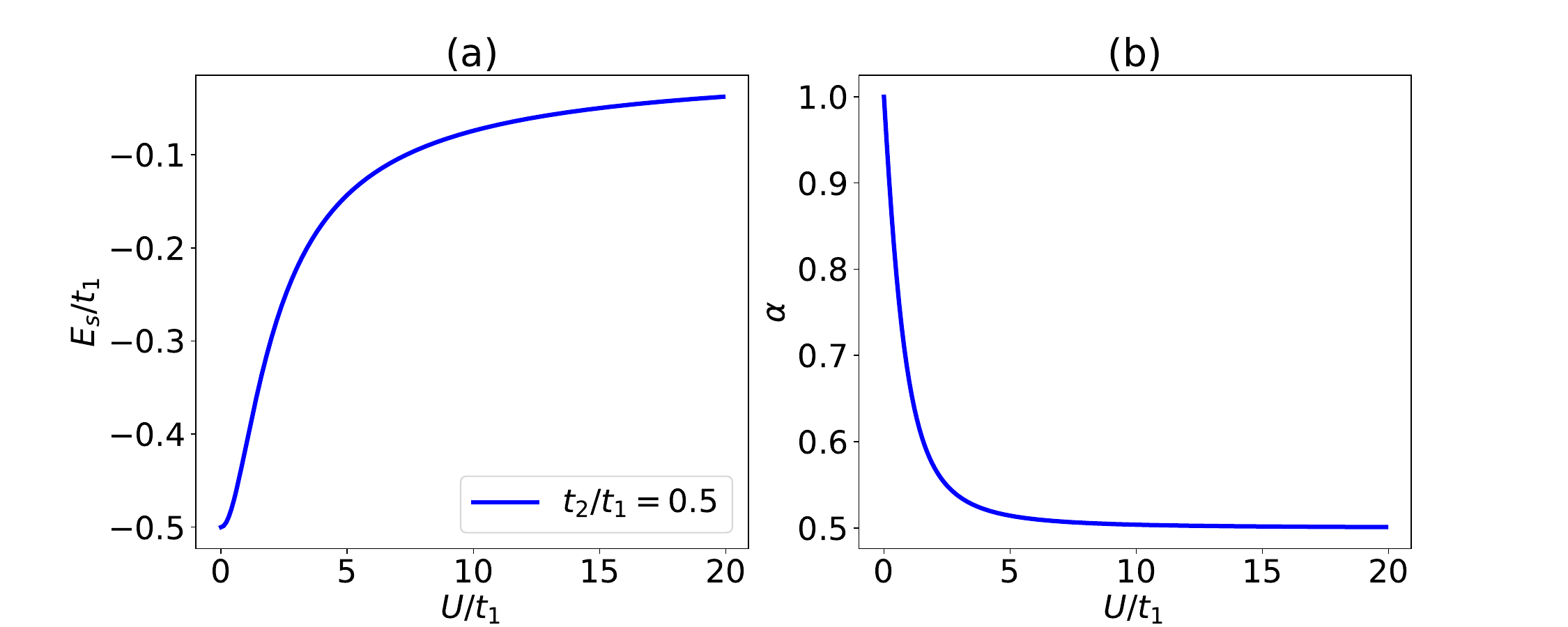}
	\caption{\label{F4}
	Graphical representation of the solutions to the equation set~\eqref{E16}. In (a), we observe the variation of the bound state energy $E_s$ (measured in units of $t$) as a function of the impurity potential $U/t$ for a single impurity located on the SSH chain with $t_2/t1=0.5$. Notably, for extremely large values of the on-site potential, the bound state energy tends to zero, which aligns with the energy of the edge states in the system. (b) illustrates the corresponding coefficient $\alpha$, as defined in the equation set~\eqref{E13}.
	}
\end{figure}

Now, let us consider the case where $U\neq 0$. We aim to demonstrate that the presence of an on-site impurity potential induces a bound state with energy $E_s$ in the gap region, $-E_g/2 < E_s < E_g/2$. To achieve this, we assume that the impurity is located on a $B$ site as depicted in Fig.~\ref{F3}. 
We begin by making an initial assumption about the form of 
the wave function associated with such an in-gap bound state as
\begin{equation}
|\psi_{\text{s}}\rangle = |\psi_A^r\rangle + |\psi_A^l\rangle + |\psi_B\rangle.
\end{equation}
Here $|\psi_A^r\rangle$ and $|\psi_A^l\rangle$ represent the contributions involving solely the A sub-lattice sites on the right and left sides, respectively, while $|\psi_B\rangle$ denotes the contribution involving solely the B sub-lattice sites and are defined  as:
\bea
|\psi_A^r\rangle&=&\sum_{i> 0} \psi_i^{A_r}  a^\dag_i|0\rangle =\sum_{i> 0} \psi^{A_r}_1 (-\alpha)^i  a^\dag_i|0\rangle, \nonumber \\
|\psi_A^l\rangle&=&\sum_{i\le 0} \psi_i^{A_l}  a^\dag_i|0\rangle =\sum_{i\le 0} \psi^{A_l}_0 (-\alpha)^i  a^\dag_i|0\rangle, \nonumber \\
|\psi_B\rangle&=&\sum_i \psi_i^B  b^\dag_i|0\rangle =\sum_i \psi^B_0 |-\alpha|^i  b^\dag_i|0\rangle, \nonumber \\ \label{E13}
\eea
where  $\psi^{{A_r}({A_l})}_j$ and $\psi^B_j$ represent the amplitudes of the wave function on the $j$th site of sub-lattice A on the right (left) side of the impurity and on sub-lattice B, respectively.
Indeed, the reason for considering two distinct parts, $|\psi_A^r\rangle$ and $|\psi_A^l\rangle$, in the case of the A sub-lattice components is due to the lack of symmetry in the wave function on the left and right sides of the impurity. The impurity introduces an asymmetry in the system, leading to different contributions from the A sub-lattice sites on each side. Therefore, to accurately describe the behavior of the wave function, we need to consider separate components for the A sub-lattice sites on the right and left sides of the impurity.
Subsequently, we insert this assumed solution into the equation 
\be
\mathcal{H} |\psi_{\text{s}}\rangle = (\mathcal{H}_{\text{SSH}} + \mathcal{H}_{\text{impurity}}) |\psi_{\text{s}}\rangle=E_s |\psi_{\text{s}}\rangle,
\ee 
enabling us to solve for the coefficients that render the equation valid.
These considerations lead to the following set of equations for the amplitudes of the wave function:
\bea
&t_1\psi_0^{A_l} + t_2\psi_1^{A_r} = (E_s - U)\psi_0^B, \nonumber \\
&(t_2 - \alpha t_1)\psi_0^B = E_s\psi_1^{A_r}, \nonumber\\
&(t_1 - \alpha t_2)\psi_1^{A_r} = -\alpha E_s\psi_0^B, \nonumber\\
&(t_1 - \alpha t_2)\psi_0^B = E_s\psi_0^{A_l}, \nonumber\\
&(t_2 - \alpha t_1)\psi_0^{A_l} = -\alpha E_s\psi_0^B. \label{SS}
\eea
Besides the three unknown amplitudes of the wave function $\psi_0^{A_l}, \psi_1^{A_r}$ and $\psi_0^B$ , we have two additional unknown quantities, namely $E_s$ and $\alpha$. To determine $E$ and $\alpha$, we need to solve the set of five equations. However, our main interest lies in finding $E_s$ and $\alpha$ directly. Therefore, we can eliminate the other three unknowns from the equations through some algebraic manipulations, resulting in the following expressions:
\begin{eqnarray}
&t_1T_1 + t_2T_2 = E_s(E_s - U) \nonumber \\
&T_1T_2 =- \alpha E_s^2. \label{E16}
\end{eqnarray}
Here, we have defined $T_1 = t_1 - \alpha t_2$ and $T_2 = t_2 - \alpha t_1$. By solving these two equations, we can obtain the desired values of $E_s$ and $\alpha$.
This set of equations has two sets of solutions: one with $E_s$ inside the gap region and the other with $E_s$ outside.
We have depicted the solutions for $E_s$ inside the gap region and its corresponding  $\alpha$ graphically in Fig.~\ref{F4}. As shown, regardless of the strength of the impurity potential $U$, a bound state emerges within the energy gap. For extremely large values of $U$ $(U\rightarrow\infty)$, the energy of the bound state approaches zero $(E_s/t_1\rightarrow 0)$, and the value of $\alpha$ converges to $\frac{t_2}{t_1}$. As we decrease the impurity strength $U$, the energy of the bound state gradually decreases until it eventually merges with the bulk states at $U=0$.

\subsubsection{Generalization of the analysis for the Kane-Mele ribbon embedded with a line defect}

\begin{figure}[t!]
  	\includegraphics[width=1.0\linewidth]{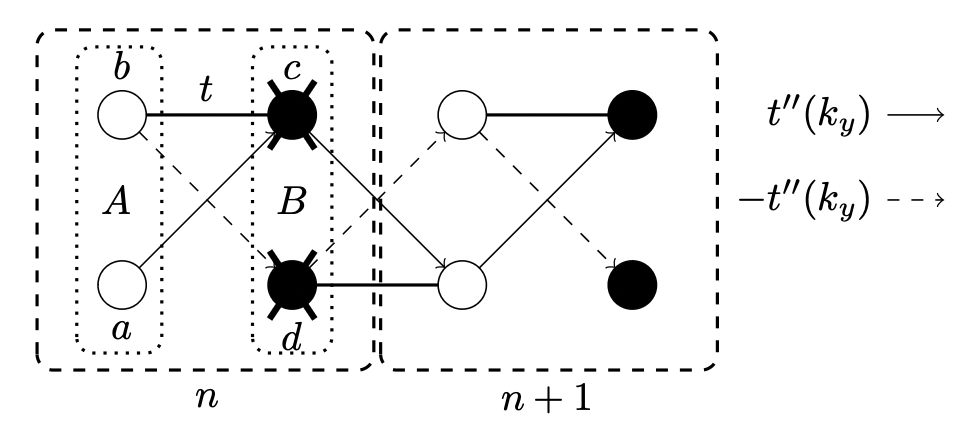}
	\caption{\label{F5}
	The figure illustrates the schematic representation of the two-leg ladder model, obtained through a dimensional reduction procedure outlined in Ref.~\cite{Amini2}, for the Kane-Mele ribbon. The unit cell of the model consists of four sites (labeled as $a, b, c,$ and $d$) denoted by a dashed rectangle, forming a plaquette. Following the analogy with the SSH chain, each plaquette is divided into two blocks, $A$ and $B$, shown by dotted lines, with each block containing two sites. The hopping parameters $t^{\prime\prime}(k_y)$ and $-t^{\prime\prime}(k_y)$ are indicated by solid and dashed vectors, respectively, and are defined as $t^{\prime\prime}(k_y) = 2\lambda_{\text{SO}}\sin\big(\dfrac{k_y}{2}\big)$.}
\end{figure}

We can extend the previous calculations to the case of a Kane-Mele ribbon with a line defect along the $y$-direction, as illustrated in Fig.~\ref{F1}. To simplify the analysis, we assume that the ribbon is wide enough along the $y$-direction so that the momentum $k_y$ can be treated as a good quantum number. This assumption allows us to employ the method and notation introduced in Ref.~\cite{Amini2}, where the Kane-Mele Hamiltonian of Eq.~\eqref{E1} is mapped onto a two-leg ladder system with a generalized SSH Hamiltonian in the $k_y$-space, $\mathcal{H}_{KM}(k_y)= \mathcal{H}_0(k_y) + \mathcal{H}_1(k_y)$. In this ladder model, the unit cell consists of a plaquette composed of four sites, denoted as $a$, $b$, $c$, and $d$. To establish an analogy with the SSH chain, we label the sites $a$ and $b$ as block $A$, and the sites $c$ and $d$ as block $B$, as depicted in Fig.~\ref{F5}. 
In this notation, the transformed Hamiltonian for the ladder system consists of two terms, $\mathcal{H}_0(k_y)$ and $\mathcal{H}_1(k_y)$. $\mathcal{H}_0(k_y)$ represents a zero-energy flat band, while $\mathcal{H}_1(k_y)$ is responsible for the dispersion of the edge band.
To proceed, we first aim to obtain the impurity bound state for the ladder system in the absence of $\mathcal{H}_1(k_y)$. A schematic representation of this ladder system is depicted in Fig.~\ref{F5}, and we can write the Hamiltonian $\mathcal{H}_0(k_y)$ as follows~\cite{Amini1}:
\begin{equation}
\mathcal{H}_0(k_y)=\sum_n [a_{n}^\dag b_{n}^\dag]\hat{t_1}\begin{bmatrix}c_{n}\\d_{n}\end{bmatrix}+ \sum_n [c_{n}^\dag d_{n}^\dag]\hat{t_2}\begin{bmatrix}a_{n+1}\\b_{n+1}\end{bmatrix}+h\cdot c,
\end{equation}
where
\begin{equation}
\hat t_1=\begin{bmatrix}
t&-2\lambda_{\text{SO}}\sin\big(\dfrac{k_y}2\big)\\
2\lambda_{\text{SO}}\sin\big(\dfrac{k_y}2\big)&0
\end{bmatrix},
\end{equation}
\begin{equation}
\hat t_2=\begin{bmatrix}
0&2\lambda_{\text{SO}}\sin\big(\dfrac{k_y}2\big)\\
-2\lambda_{\text{SO}}\sin\big(\dfrac{k_y}2\big)&t
\end{bmatrix},
\end{equation}
and the summation on $n$ runs over all the plaquettes in the ladder model.

\begin{figure}[t!]
  	\includegraphics[width=1.1\linewidth]{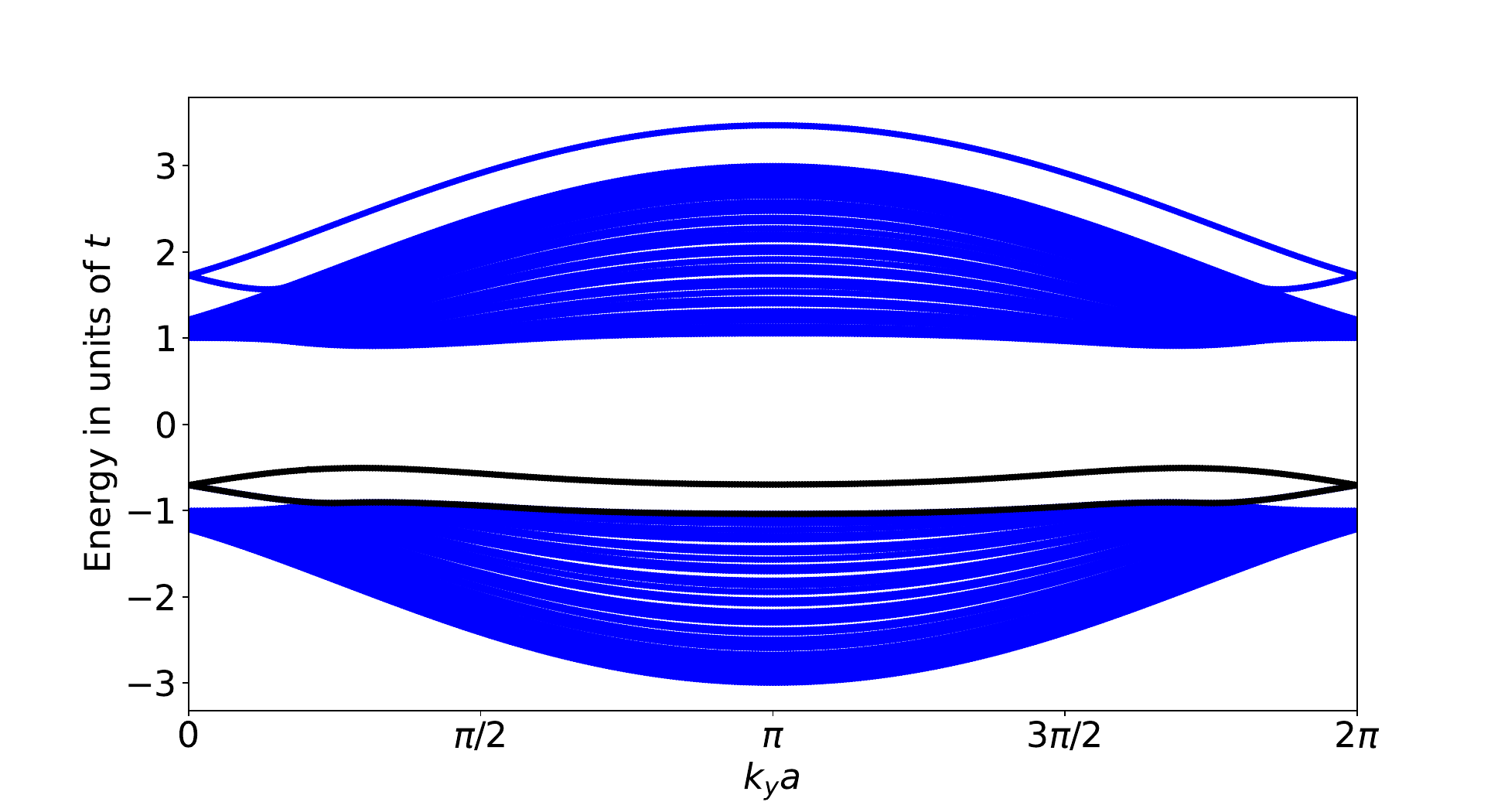}
	\caption{\label{F6}
	Electronic band structure of a Kane-Mele ribbon with zigzag boundaries, consisting of $20$ zigzag chains. The ribbon includes a zigzag chain of on-site impurities, positioned away from the edges, leading to the formation of impurity bands. The impurity bands in the gap region are highlighted in black, while the edge bands have not been shown to avoid confusion.
	}
\end{figure}

In the presence of the line defect, the Hamiltonian given by Eq.~\eqref{E2} alters the on-site potential of the corresponding block of the ladder system (based on the location of the defected zigzag chain) to the on-site energy $U$. For instance, in Fig.~\ref{F5}, the line defect is shown on block B of the $n$-th plaquette.
Thus, there exists an analogy between the SSH ladder and SSH chain, and we can derive the corresponding set of equations given in Eq.~\eqref{SS} for the ladder system by simply replacing the wave function amplitudes in the SSH chain model as two-component vectors. That is, $|\psi^{A_r}\rangle = \begin{bmatrix} \psi^{a_r}\\ \psi^{b_r} \end{bmatrix}$, $|\psi^{A_l}\rangle = \begin{bmatrix} \psi^{a_l}\\ \psi^{b_l} \end{bmatrix}$, and $|\psi^{B}\rangle = \begin{bmatrix} \psi^{c}\\ \psi^{d} \end{bmatrix}$.
This immediately results in the following set of equations:
\bea
&\hat{t_1}^\dag \begin{bmatrix} \psi^{a_l}_0\\ \psi^{b_l}_0 \end{bmatrix}+\hat{t_2} \begin{bmatrix} \psi^{a_r}_1\\ \psi^{b_r}_1 \end{bmatrix}=(E_s-U)\begin{bmatrix} \psi^{c}_0\\ \psi^{d}_0 \end{bmatrix},\nonumber\\
&(\hat t_2^\dag-\alpha\hat t_1) \begin{bmatrix} \psi^{c}_0\\ \psi^{d}_0 \end{bmatrix} =E_s \begin{bmatrix} \psi^{a_r}_1\\ \psi^{b_r}_1 \end{bmatrix},\nonumber\\
&(\hat t_1^\dag-\alpha\hat t_2)  \begin{bmatrix} \psi^{a_r}_1\\ \psi^{b_r}_1 \end{bmatrix} =- \alpha E_s\begin{bmatrix} \psi^{c}_0\\ \psi^{d}_0 \end{bmatrix},\nonumber\\
&(\hat t_1-\alpha\hat t_2^\dag)\begin{bmatrix} \psi^{c}_0\\ \psi^{d}_0 \end{bmatrix} =E_s \begin{bmatrix} \psi^{a_l}_0\\ \psi^{b_l}_0 \end{bmatrix},\nonumber\\
&(\hat t_2-\alpha\hat t_1^\dag) \begin{bmatrix} \psi^{a_l}_0\\ \psi^{b_l}_0 \end{bmatrix} =-\alpha E_s \begin{bmatrix} \psi^{c}_0\\ \psi^{d}_0 \end{bmatrix}.\label{SSN}
\eea
By the same token, this will allow us to obtain the unknown quantities $E_s(k_y)$ and $\alpha(k_y)$ through the solution of the following equations:
\begin{eqnarray}
&\hat{t}_1\hat{T}_1 + \hat{t}_2\hat{T}_2 = E_s(E_s - U) \hat{I}\nonumber \\
&\hat{T}_1\hat{T}_2 =- \alpha E_s^2 \hat{I}. \label{ES}
\end{eqnarray}
where $\hat{T}_1=\hat{t}_2^\dag-\alpha\hat{t}_1$  and $\hat{T}_2=\hat{t}_1^\dag-\alpha\hat{t}_2$.
Using the definition of $\hat{T}_1$ and $\hat{T}_2$, one can simplify the second equation in the equation set~\eqref{ES} as $\hat{T}_1\hat{T}_2=(1-\alpha)^2{t''}^2(k_y)-\alpha t^2(k_y)=-\alpha E_s^2(k_y)$.  
Indeed, the resulting energy of these bound states associated with the impurity potential $U$ explicitly depends on the wave number $k_y$. This observation indicates the formation of an impurity band~\cite{Amini3}, which leads to the creation of an impurity channel responsible for the coupling between the top and bottom edges of the ribbon.
The above-mentioned set of equations has four solutions, and two of them lie in the gap region, which we will refer to later on.
The final challenge lies in understanding the effect of $\mathcal{H}_1(k_y)$. To tackle this, we need to employ a perturbative approach following the method described in Ref.~\cite{Amini2}. This entails using the unperturbed wave functions to obtain the expectation values of $\mathcal{H}_1(k_y)$ in between. Due to the complexity of these calculations, analytical solutions are not readily available, necessitating the use of numerical computations.

To obtain more accurate results and minimize perturbation errors, we can take a different approach. We consider a zigzag ribbon of the Kane-Mele model with a line defect, represented by a zigzag chain of on-site impurities with an energy $U$. By numerically computing the band structure associated with this ribbon, we can gain deeper insights. For example, we can analyze the behavior of a ribbon with $N=20$ zigzag chains, where we set $\lambda_{\text{SO}}=0.2$ and $U=1.0$. The resulting band structure is presented in Fig.~\ref{F6}.
As previously mentioned, the presence of the line defect gives rise to two additional bands within the energy gap. In Fig.~\ref{F6}, we have highlighted these in-gap impurity bands using black colors for clarity. The edge bands are not shown in the plot to avoid confusion and better focus on the behavior of the impurity bands.

\subsection{Impurity channel analysis}

\begin{figure}[t!]
  	\includegraphics[width=1.1\linewidth]{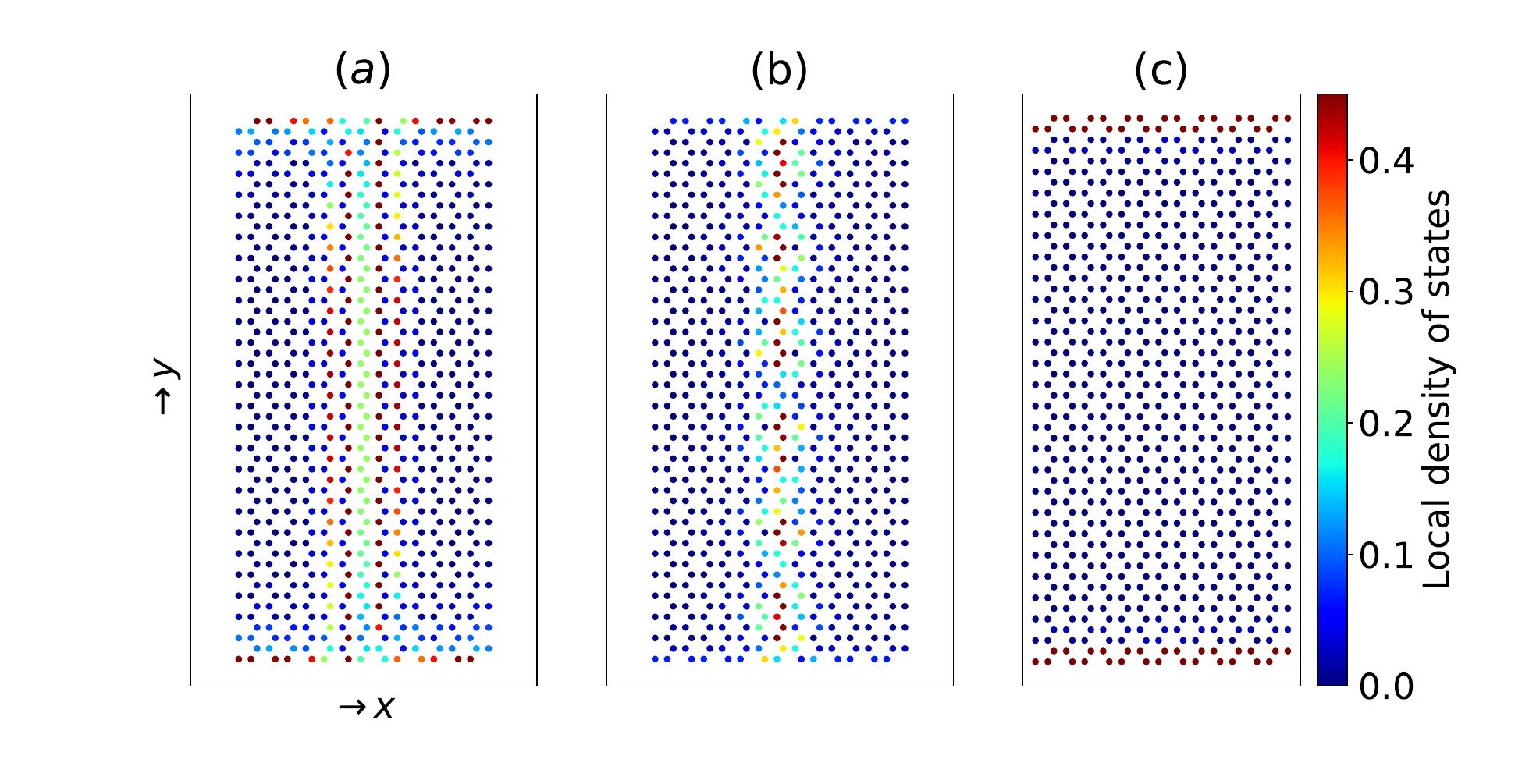}
	\caption{\label{F7}
		The figure presents spatial profiles of the local density of states on sites adjacent to the line defect at energies specified in Fig.~\ref{F2}. The scattering region is a Kane-Mele ribbon (depicted in Fig.~\ref{F1}) with a width of $W=26$ and spin-orbit coupling $\lambda_{\text{SO}}=0.2 t$ that hosts a line defect with on-site potential strength $U=1.0 t$. 
(a) shows the LDOS at energy $E/t=-0.88$ with complete suppression of the transmission coefficient. 
(b) shows the LDOS at energy $E/t=-0.68$ with partial suppression of the transmission coefficient.
(c) displays the LDOS at $E/t=0.0$ with an unchanged transmission coefficient.	
}
\end{figure}

Another approach to comprehend the formation of the impurity channel is by investigating the LDOS at energies where topological protection is broken. By analyzing the LDOS, which is related to the matrix elements of the retarded Green's function as $\rho_i(E)=-\frac{1}{\pi} \text{Im} [\langle i |G_r(E) | i\rangle]$, we can gain insights into how the presence of the impurity potential affects the electronic states in real space and leads to the emergence of the impurity channel.
Fig.~\ref{F7} displays the spatial distribution of the LDOS for the scattering region depicted in Fig.~\ref{F1} near the line defects at energies specified in Fig.~\ref{F2}. We are interested in the energies near the lower gap edge where the presence of line defects has a visible impact. 
As before, we consider the parameters $\lambda_{\text{SO}}=0.2 t$ and $U=1.0 t$ for our calculations. These values remain consistent with our previous analysis and investigations of the system's behavior in the presence of the line defect.
Specifically, Fig.~\ref{F7}(a) illustrates the LDOS map at a specific energy value of $E/t=-0.88$, showcasing a complete suppression of transmission, indicating the formation of the impurity channel. 
Additionally, Fig.~\ref{F7}(b) depicts the LDOS at $E/t=-0.68$, where the transmission is not entirely suppressed. However, due to the presence of impunity channel connecting the top and bottom edges of the ribbon partial backscattering existence.
For comparison, Fig.~\ref{F7}(c) shows the LDOS at $E/t=0.0$, which exclusively displays only the presence of edge states (without formation of any impurity channel), as anticipated.
These results provide valuable insights into the impact of the line defect on the edge states' transmission and the formation of the impurity channel in the system.

\section{CONCLUDING REMARKS \label{Sec.IV}}
In conclusion, we have investigated the impact of a line defect in the form of on-site impurities, arranged in a zigzag chain connecting the top and bottom edges of a Kane-Mele armchair ribbon. Our study focused on the scattering of edge states by this line defect and its effect on the topological protection of these states.
We found that the presence of the line defect can significantly break the topological protection of the edge states, leading to backscattering of the helical electrons. As a consequence, the conductance of the system becomes lower than that of the pristine ribbon. Our analysis involved a combination of numerical and analytical calculations to elucidate the underlying mechanism behind this backscattering phenomenon.
The key result of our study is the observation that the line defect induces an impurity channel that spans the width of the ribbon. This impurity channel facilitates the backscattering of helical electrons, causing a suppression of the quantized transmission within specific energy windows.
In summary, our work sheds light on the role of line defects in disrupting the topological protection of edge states in the Kane-Mele model. These findings have implications for the design and understanding of topological systems and their robustness against imperfections and defects.

\begin{acknowledgments}
We extend our sincere gratitude to the office of graduate studies at the University of Isfahan(UI) for their generous support and for providing us with research facilities, which were instrumental in carrying out this study. Additionally, MA would like to acknowledge the support received from the Abdus Salam (ICTP) associateship program.
\end{acknowledgments}


\end{document}